# Bulk and surface states carried supercurrent in ballistic Nb-Dirac semimetal Cd$_3$As$_2$ nanowire-Nb junctions

Cai-Zhen Li,[1,†] Chuan Li,[2,†] Li-Xian Wang,[1,†] Shuo Wang,[1] Zhi-Min Liao,[1,3,*] Alexander Brinkman,[2,*] Da-Peng Yu[1,4]

[1] State Key Laboratory for Mesoscopic Physics, School of Physics, Peking University, Beijing 100871, China.

[2] MESA+ Institute for Nanotechnology, University of Twente, 7500 AE Enschede, The Netherlands.

[3] Collaborative Innovation Center of Quantum Matter, Beijing 100871, China.

[4] Institute for Quantum Science and Engineering and Department of Physics, South University of Science and Technology of China, Shenzhen 518055, China

†These authors contributed equally to this work.

* liaozm@pku.edu.cn; a.brinkman@utwente.nl

**A three-dimensional Dirac semimetal has bulk Dirac cones in all three momentum directions and Fermi arc-like surface states, and can be converted into a Weyl semimetal by breaking time-reversal symmetry. However, the highly conductive bulk state usually hides the electronic transport from the surface state in Dirac semimetal. Here, we demonstrate the supercurrent carried by bulk and surface states in Nb-Cd$_3$As$_2$ nanowire-Nb short and long junctions, respectively. For the ~1 μm long junction, the Fabry-Pérot interferences induced oscillations of the critical supercurrent are observed, suggesting the ballistic transport of the surface states carried supercurrent, where the bulk states are decoherent and the topologically protected surface states still keep coherent. Moreover, a superconducting dome is observed in the long junction, which is attributed to the enhanced dephasing from the interaction between surface and bulk states as tuning gate voltage to increase the carrier density. The superconductivity of topological semimetal nanowire is promising for braiding of Majorana fermions toward topological quantum computing.**

Dirac semimetals are newly emerging three-dimensional (3D) topological materials that possess gapless Dirac dispersions in bulk, protected by time-reversal symmetry and inversion symmetry [1,2]. $Cd_3As_2$ has been identified to be a 3D Dirac semimetal by angle-resolved photoemission spectroscopy (ARPES) [3,4] and scanning tunneling microscopy (STM) [5] experiments. Owing to the unusual band dispersions, $Cd_3As_2$ exhibits many exotic transport phenomena, such as ultra-high carrier mobility [6], the chiral anomaly effect [7] and Fermi-arc like surface states related quantum oscillations [8-10]. The combination of topological material and superconductor is one of the promising routes towards topological superconductivity [11-14]. A topological superconductor is a crucial state of matter associated with Majorana fermions [15,16], which obey non-Abelian statistics and can provide quantum states that are topologically protected from decoherence. Recently, theoretical investigations have proposed that the Dirac semimetal $Cd_3As_2$ is promising for topological superconductivity [17,18]. In addition, the signatures of unconventional superconductivity in $Cd_3As_2$ have been reported through pressure-loaded [19] and point-contacted [20,21] experiments. However, the proximity effect induced superconductivity in $Cd_3As_2$ has still not yet been achieved, which is more important for practical devices without complicated pressure loading. On the other hand, the transport properties of the surface states of Dirac semimetal are difficult to be manifested due to the inevitable bulk conductance.

Here, we report the gate voltage, magnetic field, microwave irradiation and temperature modulated supercurrent in $Nb/Cd_3As_2$ nanowire/Nb hybrid structures, taking advantages of the high crystal quality, the large surface to volume ratio, and the feasibility of nanoscale device fabrication of $Cd_3As_2$ nanowires. The surface states carried supercurrent is observed in a junction with ~1μm channel length, where the bulk states are decoherent and the topologically protected surface states still keep coherent.

The $Cd_3As_2$ nanowires were synthesized by chemical vapor deposition method [7]. The transmission electron microscopy (TEM) image shown in Fig. 1a indicates the single crystalline nature of the nanowire. Individual nanowires were transferred to Si substrates with 285 nm thick oxide layer. After the process of standard electron beam lithography, Nb (100nm)/Pd (5nm) electrodes were deposited by magnetron sputtering. The thin Pd layer was deposited with the aim to prevent Nb from oxidization. To improve the interface quality, the nanowire surface was etched with Ar-ion sputtering *in situ* before Nb deposition. The scanning electron microscopy (SEM) image in Fig. 1b shows the nanowire-Nb junctions. The nanowire diameter is ~100 nm. Three junctions with channel length of 100 nm (Junction-A), 118 nm (Junction-B), and 1 μm (Junction-L) were measured, respectively.

Electrical measurements were carried out in an Oxford Instrument Triton Cryofree dilution refrigerator with a base temperature of 12 mK. The heavily doped Si substrate and the $SiO_2$ dielectric together served as a back gate to tune the Fermi level of the nanowire. Figure 1c shows the current-voltage (*I-V*) curves from Junction-A measured at gate voltages $V_g$ = 16, 0, -16 V, respectively. A clear zero resistance state and a dissipative state above the critical switching current $I_c$ are observed. The $I_c$ has a close relationship with $V_g$. Figure 1d displays the mapping of the differential resistance d$V$/d$I$ with $V_g$ and source-drain current $I_{sd}$. The $I_{sd}$ was swept from negative to positive and the upper boundary of the purple region (the superconducting state) corresponds to $I_c$. The $I_c$ decreases monotonically when sweeping $V_g$ to lower values. For $V_g$ < -25 V, $I_c$ saturates to a small but finite value. The normal state resistance $R_n$ is extracted at $I_{sd}$ = 400 nA. The $I_cR_n$ product is nearly constant as function of gate voltage with a mean value of 115 μV (Supplementary Fig. S1), indicating that the Josephson device is in the short junction limit where the electrode separation is less than the coherence length of the interlayer. $Cd_3As_2$ nanowire devices fabricated using a similar process but with a long channel length (~ μm) and with Au contacts usually show the Dirac point near 0 V. However, the Dirac point of the short Nb-$Cd_3As_2$-Nb junction is still not reached at $V_g$ = -80 V (Supplementary Fig. S2). This result suggests that the Nb contacts provide heavy electron doping into the $Cd_3As_2$ nanowire.

The $I_c$ demonstrates a monotonous decay with increasing the perpendicular magnetic field *B* (Fig. 2a). Similar behavior has also been observed in InSb nanowire based Josephson junctions [22]. The magnetic field dependence of $I_c$ agrees well with a Gaussian fitting (Fig. 2b), which is usually observed in narrow Josephson junctions as the junction width (*W*) is comparable or smaller than the magnetic length $\xi_B = \sqrt{\Phi_0/B}$, where $\Phi_0$ corresponds to the magnetic flux quantum [23-25]. The characteristic magnetic field can be expressed as $B_c = \Phi_0/S$, where *S* is the junction area perpendicular to the magnetic field. For Junction-A, the calculated $B_c$ is as large as 0.2 T, which is well consistent with the experimental observations. For $B < B_c$, the magnetic length $\xi_B$ is smaller than the nanowire diameter ~100 nm, and the narrow junction model applies. The mapping of d$V$/d$I$ as a function of $I_{sd}$ and *rf* power at 7 GHz was shown in Fig. 2c. The Shapiro steps are observed at $V = nhf/(2e)$ (Fig. 2d), where *h* is Plank's constant, *f* is the microwave frequency, and *n* is an integer. The odd as well as even steps are observed up to $n = 5$, revealing a dominant $2\pi$ periodic contribution to the current-phase relation. Although the topological superconductivity would lead to a $4\pi$-periodic Josephson supercurrent, its absence in experiments is also understandable, because the supercurrent amplitude of the $4\pi$-period is much smaller than that of the $2\pi$-period. Moreover, the $4\pi$-periodic supercurrent is sensitive to the quasi-particle scattering and the transparency of the interface [26,27]. Further optimizing the device interface and the measurement conditions would help to reveal the $4\pi$-periodic supercurrent.

To reduce the influence of electron-doping from the Nb contacts, a long junction (Junction-L) with a channel length ~1μm was studied. Figure 3a displays the $V_g$ dependence of d$V$/d$I$ measured at an excitation current $I_{ac} = 2$ nA. Clearly, the long junction still demonstrates supercurrent in the $V_g$ region of 0 to 20 V. While tuning $V_g$ below 0 V, the resistance increases dramatically and the supercurrent disappears. Also for $V_g > 20$ V, a non-zero resistance state emerges, showing dissipative transport. The gate switchable supercurrent indicates that there are different conduction channels depending on the Fermi level of the nanowire. The critical current is ~10 nA at $V_g = 4$ V, while almost decreases to 0 at $V_g = 20$ V (Fig. 3b). The overall evolution of d$V$/d$I$ with $V_g$ and $I_{sd}$ is exhibited in Fig. 3c. The $I_c$ has its maximum at around $V_g = 4$ V.

Such a convex behavior of $I_c$ as a function of gate voltage is very different from the concave behavior as observed in graphene [28,29] and InAs nanowire [30] based Josephson junctions. Generally, the $I_c$ should increase with increasing the electron concentration, such as the behavior in 100 nm short junction (Fig. 1d). Interestingly, Fig. 3c shows that the $I_c$ decreases with increasing $V_g$ away from 4 V, reminding one of the superconductivity dome phenomena [31-35]. The dome-shaped superconductivity is not restricted in oxide compounds, such as Na-doped $WO_3$ [31], $KTaO_3$ [32], $LaAlO_3/SiTiO_3$ interface [33], but also occurs in a band insulator $MoS_2$ [34]. However, the mechanism behind the superconductivity dome is still under debate. It is attributed to the phonon softening from a structural transition in Na-doped $WO_3$, and the quantum critical point in $KTaO_3$ and $LaAlO_3/SiTiO_3$ interface. While, Jianting Ye *et al.* believed that there is a more universal and non-material related mechanism for the superconducting dome [31]. Considering the existence of both bulk and surface states in the Dirac semimetal nanowires, we proposed a mechanism related to the bulk-surface states interaction induced dephasing for the observed superconducting dome. Liao *et. al* [36] have reported the enhanced electron dephasing due to the coupling between surface states and charge puddles in the bulk in a 3D topological insulator. In Dirac semimetals, there is a highly conductive bulk state, and the scatterings between surface and bulk states would produce significant dephasing. As tuning the gate voltage to increase the electron density, the interaction between bulk and surface states will be enhanced. It destroys the coherent transport of the Cooper pairs and the supercurrent.

Unambiguous oscillations of $I_c$ as function of gate voltage are observed. The rich pattern of $dV/dI$ as a function of $I_{sd}$ and $V_g$ in the region of $-1 < V_g < 0$ V is shown in Fig. 3d. The oscillation may be attributed to a Fabry-Pérot (F-P) interference [37-39]. Due to the electron-doping effect near the Nb contacts, *n-p* walls may be formed near the contacts (Fig. 3e) as the central part of the channel is tuned into the hole dominant region by gate voltage. The barrier and reduced transmission probability near the contacts lead to the partial reflection of carriers and the formation of a F-P cavity. The modulation of the critical current shows a periodic dependence with $k_F$ (Supplementary Fig. S3) and the corresponding fast Fourier transform (FFT) analysis (Fig. 3f) gives an oscillation period of $\Delta k_f \sim 4.01 \, \mu m^{-1}$. An effective cavity length $L_{\text{eff}} \sim 780 \, nm$ is then obtained from the F-P resonance condition, i.e. $k_f L_{\text{eff}} = n\pi$ (here we take $n = 1$). The $L_{\text{eff}}$ is slightly shorter than the channel length ~ 1 μm, which is consistent with the existence of electron-doped regions near Nb contacts. The F-P interferences also result in pronounced oscillations of the normal state conductance, $G_n$, as a function of both $V_g$ and $V_{sd}$ under large negative gate voltages (Supplementary Fig. S4). Remarkably, the oscillations of $I_c$ with $V_g$ do not match with the oscillations in $G_n$ (Supplementary Fig. S5), in contrast to observations in ballistic graphene Josephson junction [37,38].

The observation of the F-P interference of the critical current in a micrometer-long $Cd_3As_2$ nanowire channel indicates that the supercurrent is of ballistic nature, i.e. the electrode separation is smaller than the elastic mean free path. Given the fact that the mean free path of bulk carriers in our samples is much smaller than the channel length ~1 μm (Supplementary Fig. S6), we can likely conclude that the supercurrent in the long junction is mainly carried by the surface states. The surface states may have a higher mobility due to the very different topology of the Fermi surface. The contribution of the surface states to transport in similar $Cd_3As_2$ nanowire was previously revealed by Aharonov-Bohm oscillations [8], benefitting from the large surface-to-volume ratio. Moreover, when the Fermi level is located near the Dirac point, the bulk conductions can be significantly suppressed due to the vanishing DOS at the Dirac point. The different oscillations in $I_c$ and $G_n$ may be attributed to the different conduction channels between the supercurrent and the normal conductance of the nanowire, where the supercurrent may be carried by surface states and the normal conductance is carried by both surface and bulk states. The surface states in a Dirac semimetal consist of two Fermi arcs, which connect the two surface projections of the bulk's Dirac points. For Junction-L, the increased electron carriers by tuning gate voltage not only enlarge the electron pockets, but also enhance the interaction between bulk and surface states, possibly also explaining the decrease of $I_c$ as function of gate voltage.

To verify the ballistic nature of the Josephson supercurrent, the temperature dependence of $I_c$ at different gate voltages was investigated (Fig. 4). By applying a gate voltage, it is possible to tune the transmission coefficient as well as the chemical potential and the ratio of surface/bulk contributions to transport. For Junction-A, The $I_c(T)$ dependence at $V_g$ = -80 and -50 V show a linear behavior (Fig. 4a). At $V_g$ = 0 V, the $I_c(T)$ dependence shows a concave behavior at high temperatures, and gradually saturates at low temperatures. In the short and ballistic limit, the critical supercurrent is given by [29,40]:

$$I_c(\theta,\tau,T) = \max_\theta \left( \alpha \times \frac{\Delta}{eR_n} \times \frac{\sin\theta}{\sqrt{1-\tau\sin^2\theta/2}} \times \tanh\left(\frac{\Delta}{2k_BT}\sqrt{1-\tau\sin^2\theta/2}\right) \right), \quad (1)$$

where $\tau$ is the transmission coefficient of the SN interface, $\theta$ is the phase difference between the two superconducting electrodes, $R_n$ is the normal resistance and $\alpha$ corresponds to the reduction of $I_c$ due to a non-ideal environment. The temperature dependent superconducting gap is assumed to be $\Delta \approx \Delta_0\sqrt{1-(\frac{T}{T_c})^2}$, where $\Delta_0$ is the gap as T → 0, and $T_c$ is the critical temperature. As the multiple Andreev reflections in our measurements are not distinct enough to determine $\Delta_0$, it was treated as a fitting parameter. At $V_g$ = 0 V, the fitting results give $\tau$ = 0.37, $\alpha$ = 0.40, and $\Delta_0$ = 0.36 meV. For $V_g$ = -80 V, the fitting results give $\tau$ = 0.23, $\alpha$ = 0.32, and $\Delta_0$ = 0.14 meV. The reduced transmission coefficient $\tau$ at -80 V is consistent with the observed Nb contact induced *n*-type doping near the $Cd_3As_2$/Nb interface. The reduced gap $\Delta_0$ is consistent with the decrease of the $I_cR_n$ product. Since the gate also influences the bulk contribution to $R_n$, especially when $V_g$ approaches the Dirac point, the overall magnitude of $I_c$ is changing with gate too. In the long junction-L, the $I_c(T)$ dependence shows that $I_c$ at $V_g$ = 4 V is always larger than at $V_g$ = 10 V (Fig. 4b), which would be consistent with the fact that the surface carried supercurrent can be suppressed by the hybridization between surface and bulk states [13]. The observation of the F-P oscillations indicates the supercurrent in Junction-L may also flow *via* ballistic transport, and the equ. (1) should be applicable to analyze the $I_c(T)$ dependence. As shown in Fig. 4b, the fittings are in good agreement with the experimental data. The fitted superconducting gap $\Delta_0$ is about 0.086 and 0.118 meV under $V_g$ =10 and 4 V, respectively. Given that $\Delta_0 \approx 1.76 k_B T_c$, the corresponding critical temperature $T_c$ at $V_g$ =10 and 4 V is 0.57 and 0.78 K, respectively, which is consistent with the extrapolated value from the $I_c(T)$ dependence. Such a convex saturating behavior of $I_c(T)$ dependence is similar to that observed in monolayer graphene vertical Josephson junction in the ballistic short junction limit [40]. In the long junction limit [29], the $I_c(T)$ dependence should have obeyed the exponential scaling $I_c \propto \exp(-k_B T/\delta E)$, where $\delta E \approx \hbar v_f / 2\pi L$. The reason that our longest junction still can be fitted by short junction theory is the combination of a large Fermi velocity and small induced gap value, resulting in a large coherence length of the order of the device length (Supplementary Fig. S7).

In summary, the proximity effect induced supercurrent has been realized in Dirac semimetal $Cd_3As_2$ nanowire based Josephson junctions. The superconductivity in Weyl/Dirac semimetals with nontrivial band structure of bulk Dirac/Weyl nodes, as well as the Fermi arc surface states, is promising for the presence of Majorana bound states. This makes the combination of superconductivity and Weyl/Dirac semimetal attractive, opening an avenue for the search of topological superconductivity and Majorana fermions. Plainly, the realization of Josephson supercurrent through the surface state is an important step towards the observation of unconventional superconductors and Majorana bound states in such devices.

**Acknowledgments**

We thank Hans Hilgenkamp for discussions and supports. This work has been supported by National Key Research and Development Program of China (Nos. 2016YFA0300802) and NSFC (No. 11774004).

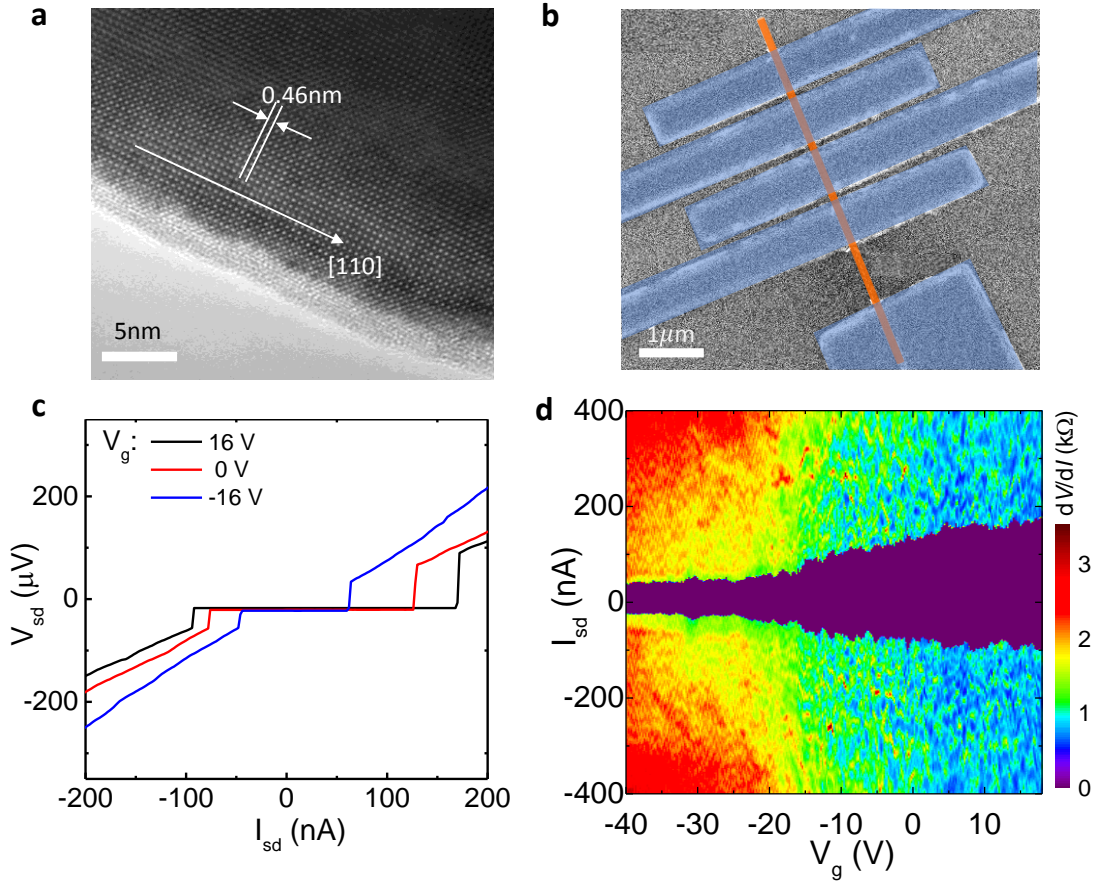

**Figure 1 | Characterization of Nb/Cd$_3$As$_2$ nanowire/Nb junction.** (a) High-resolution TEM image of a typical Cd$_3$As$_2$ nanowire, showing the [110] growth direction. (b) SEM image (false color) of the Nb/Cd$_3$As$_2$ nanowire/Nb Josephson junctions. The diameter of the nanowire is 100 nm. The inter-electrodes separation lengths are 136 nm, 118 nm (Junction-B), 100 nm (Junction-A) and 1 μm (Junction L) (from left-top to right-bottom). (c) The *I-V* characteristics of Junction-A at different gate voltages. (d) Differential resistance d$V$/d$I$ on a color scale as a function of source-drain current $I_{sd}$ and gate voltage $V_g$ of Junction-A. The central purple area corresponds to the superconducting region. The $V_g$ dependence of the critical current $I_c$ can be identified from the upper edge of the purple area. The measurements in (c) and (d) were performed at the base temperature of 12 mK.

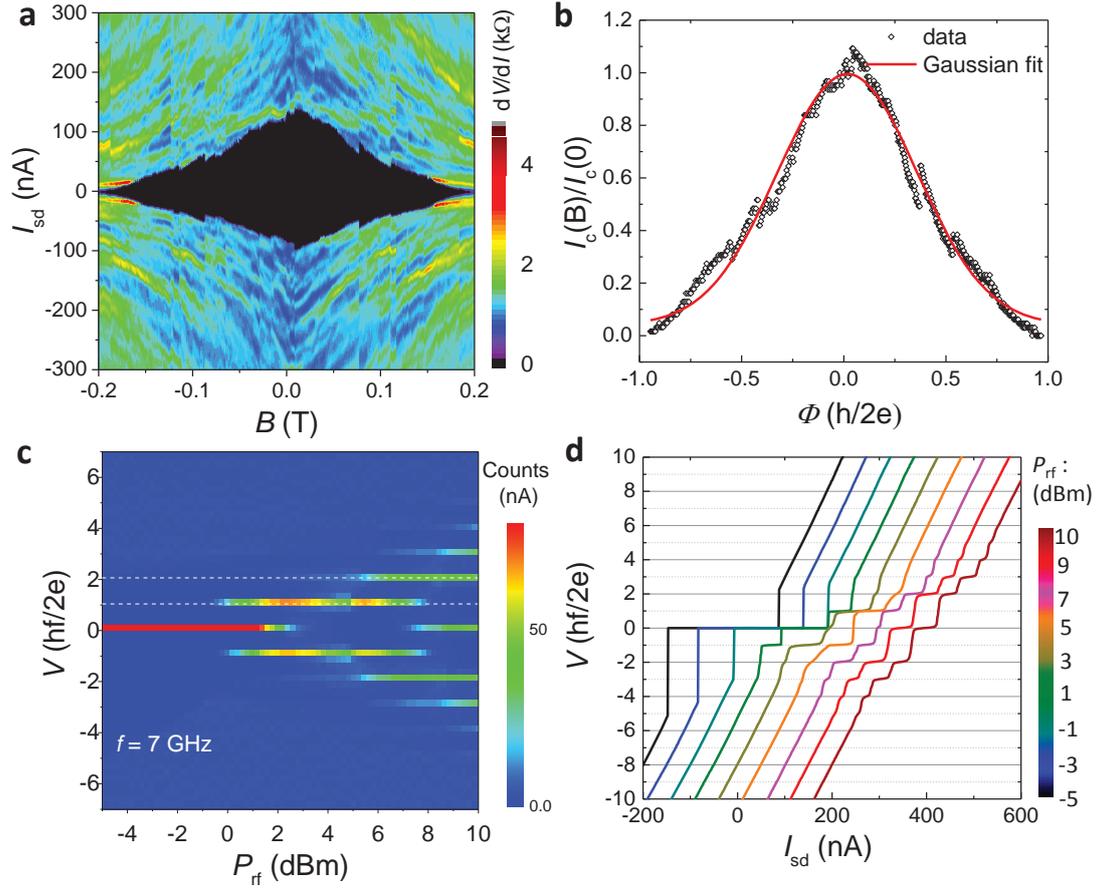

**Figure 2 | Josephson effects under perpendicular magnetic field and microwave irradiation**. **(a)** Differential resistance d$V$/d$I$ on a color scale as a function of magnetic field $B$ and source-drain current $I_{sd}$ for Junction-A. **(b)** Normalized critical current $I_c(B)/I_c(0)$ versus $\Phi$. The red line represents the Gaussian fit. **(c)** $I_{sd}$ plotted on a color scale as a function of source-drain voltage $V$ and microwave excitation power $P_{rf}$ of Junction-B (channel length 118 nm) at the $rf$ frequency of 7 GHz. **(d)** I-V characteristics of the Junction-B under the 7 GHz $rf$ excitation with different power values. The curves are shifted for clarity. Clear Shapiro steps are observed. The measurements were performed at the base temperature of 12 mK.

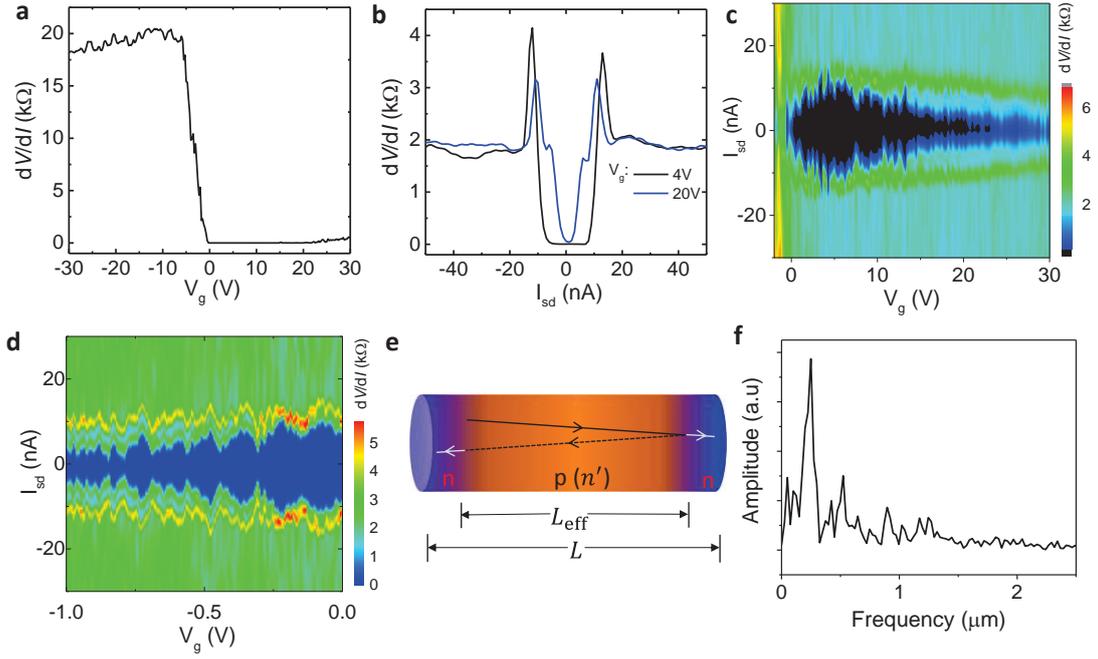

**Figure 3 | The Josephson supercurrent of Junction-L (channel length 1μm). (a)** Differential resistance as a function of gate voltage at $I_{sd}$ = 0 nA with an excitation current $I_{ac}$ = 2 nA. The supercurrent exists in the $V_g$ region of 0 ~ 20 V. While at negative gate voltages, the resistance increases sharply. **(b)** Differential resistance d$V$/d$I$ as a function of source-drain current $I_{sd}$ at $V_g$ = 4 and 20 V. **(c)** Differential resistance d$V$/d$I$ on a color scale as a function of $I_{sd}$ and $V_g$. The central dark area corresponds to the superconducting region. **(d)** Elaborate measurements of d$V$/d$I$ as a function of $I_{sd}$ and $V_g$ in the region of -1 ~ 0 V. $I_c$ oscillations are observed clearly. **(e)** Sketch of the Fabry-Pérot resonator in the nanowire between two Nb contacts. **(f)** FFT analysis of the $I_c(k_f)$ oscillation. The FFT peak F = 0.249 corresponds to an oscillation period $\Delta k_f \sim 4.01\ \mu m^{-1}$.

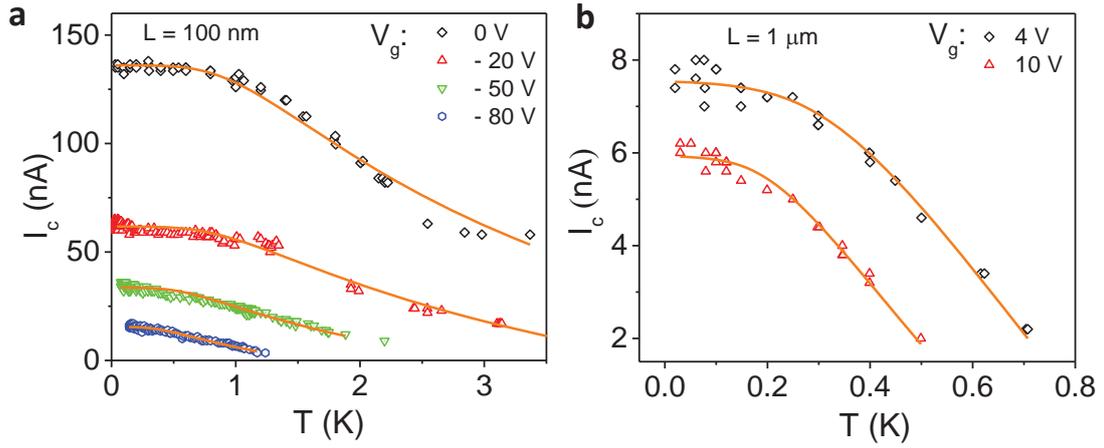

**Figure 4 | Temperature dependence of the critical current. (a)** The data for junction-A at $V_g = 0$, -20, -50 and -80 V. **(b)** $I_c(T)$ for junction-L at $V_g = 4$ and 10 V. The solid curves are the fitting results according to the ballistic short junction regime.

# Supplementary Materials

## 1. The $I_c R_n$ product of Junction-A.

For Junction-A, the critical current $I_c$, normal state resistance $R_n$ and the $I_c R_n$ product as a function of gate voltage are shown in Fig. S1. The normal state resistance $R_n$ is deduced from the differential resistance at $I_{sd} = 400 nA$. The $I_c R_n$ product is nearly a constant as $V_g < -25V$, while it fluctuates strongly with a nearly constant background as $V_g > -25V$. The maximum of $I_c R_n$ (248 µV) yields a lower limit on the induced gap $I_c R_n \leq \frac{\Delta_i}{e}$. The coherent length $\xi = \frac{\hbar v_f}{\pi \Delta_i}$ (1, 2) thus can be estimated to be between 254 to 845 nm, according to the Fermi velocity in the range of $v_f \simeq 3 \times 10^5 - 1 \times 10^6 m/s$ (3, 4). Consequently, the transport in Junction-A should be close to the short junction limit.

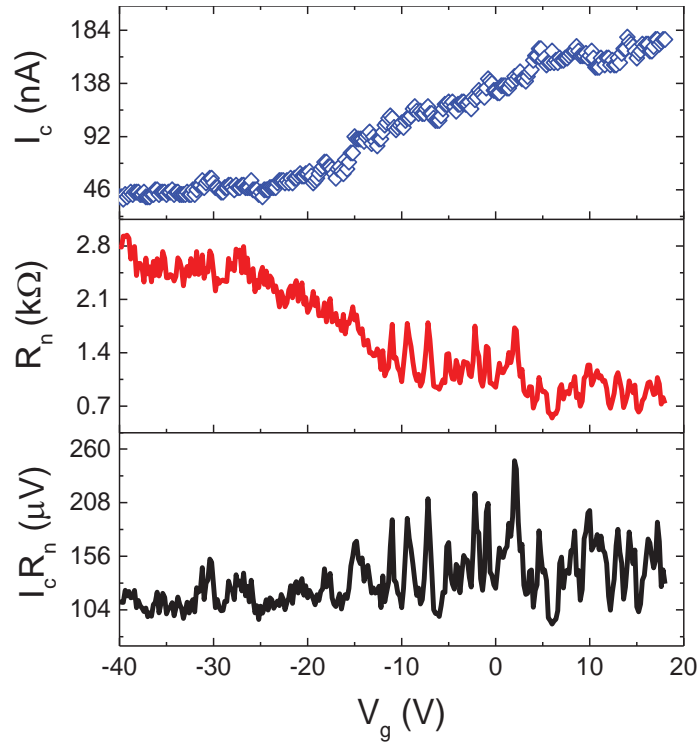

**Supplementary Fig. S1.** The critical current, normal state resistance and their product $I_c R_n$ in Junction-A as a function of gate voltage.

## 2. Electron doping effect due to Nb contacts and the gating effect in $Cd_3As_2$ nanowire.

Figures S2(A, B) show the transfer curves of individual $Cd_3As_2$ nanowire with Nb and Au contacts, respectively. The device with Nb contacts was measured at 10 K with a channel length ~118 nm, and the Dirac point is not observed even at $V_g = -80V$. For comparison, the transfer curve of the nanowire with Au contacts measured at 1.5 K with a channel length ~1.6 μm was displayed in Fig. S2B. In contrast, the Dirac point of the device with Au contacts is near $V_g = 0V$, indicating that the Fermi level is very close to the Dirac point. The long channel length of the device with Au contacts makes the center part of the nanowire far from the influences of Au contacts, thus revealing the intrinsic property of the nanowire. Because the temperature difference could not result in such a big difference for the location of Dirac point, this comparison between junctions with Nb and Au contacts indicates the heavy electron-doping of the $Cd_3As_2$ nanowire from the Nb contacts. Similar doping effect was also observed in graphene-based Josephson junctions (*5, 6*).

For the gate modulation effect, the carrier concentration of the nanowire is obtained by $n = \frac{C}{l}\frac{1}{eS}(V_g - V_D)$, where $\frac{C}{l} = \frac{\pi\varepsilon_0\varepsilon_r}{\cosh^{-1}(\frac{r+h}{r})}$ (*7*), $C$ is the capacitance of the oxide layer ($SiO_2$), $\varepsilon_r = 3.9$ is the relative dielectric constant of $SiO_2$, $h = 285$ nm is the $SiO_2$ thickness, $r$, $l$, and S are the radius, length, and cross-section area of the nanowire, respectively. When the Fermi level is near the Dirac point, the carriers can be modulated uniformly in the nanowire. For example, the transfer curve in Fig. S2B shows that the Dirac point $V_D$ is ~2 V. The resistance decreases sharply when tuning the gate voltage positively ($2 < V_g < 10\ V$) or negatively ($-30 < V_g < 2V$), a clear indication of the gating effect. Considering the nanowire diameter $d = 100$ nm and channel length $L = 1.6$ μm, the carrier density induced by a gate voltage $V_g$=10 V is estimated to be $n \sim 2.67 \times 10^{17} cm^{-3}$, and the corresponding Fermi wave-vector $k_f \sim 0.02/Å$. Based on the linear dispersion of Dirac semimetal, the rise of the Fermi energy $E_F = \hbar v_f k_f$ is around 39 ~ 131 meV by considering the Fermi velocity $v_f$ in the range of $0.3\sim1 \times 10^6 m/s$. While for $V_g > 15V$, the resistance is almost invariant, indicating that the gating effect is ineffective due to the screening effect, as the nanowire is with high carrier density.

For Junction-A (Fig. S2A), the carriers are gradually reduced as tuning the gate voltage from $V_g$ = 30 to -80 V, leading to the increase of resistance. Although the gating capability in Junction-A may be weakened compared with that close to the Dirac point, the electrons can still be depleted by the negatively biased gate voltage even that the nanowire is heavily *n*-doped by Nb contacts.

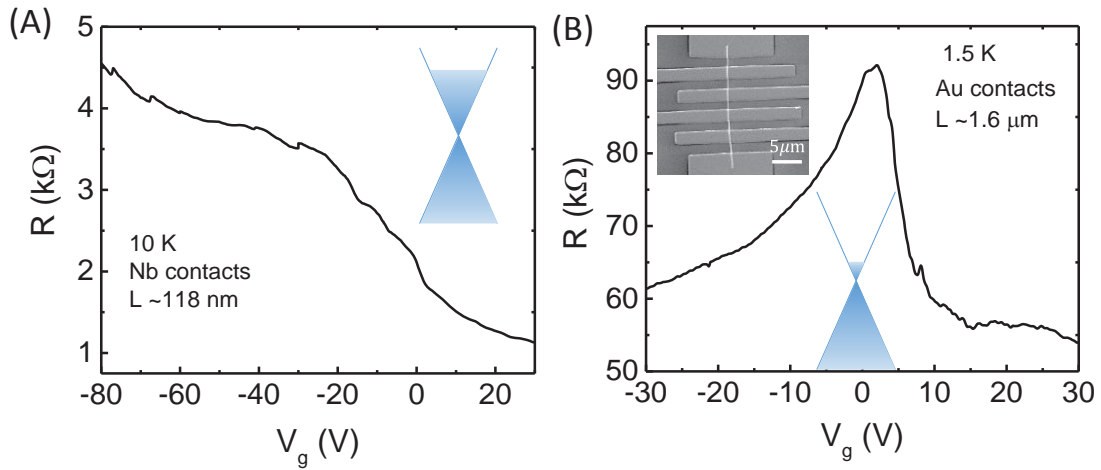

**Supplementary Fig. S2.** **(A)** Gate voltage dependence of resistance of individual Cd$_3$As$_2$ nanowire with Nb contacts measured at 10 K. The channel length is ~118 nm. Inset: Schematic diagram of the Dirac cone shows the high Fermi level with highly electron doping. **(B)** Gate voltage dependence of resistance of Cd$_3$As$_2$ nanowire with Au contacts measured at 1.5 K. The channel length is ~1.6 μm. Top inset: SEM image of the device with Au contacts. Bottom inset: Schematic diagram of the Dirac cone shows the Fermi level close to the Dirac point.

## 3. The $I_c(k_F)$ oscillations of Junction-L.

According to the transfer curve in Fig. S6, the Dirac point is roughly estimated to be $V_D = -1.5V$, where the p-n junction is starting to be forming and the resistance changes most sharply. Therefore, the $I_c \sim V_g$ dependence can be converted to $I_c \sim k_f$ dependence according to the gate induced carrier density and the band structures of $Cd_3As_2$ near the Dirac point. As shown in Fig. S3, the $\Delta I_c$ shows a periodic oscillation as a function of $k_f$ in the range of $k_f \sim 95 - 115 \; \mu m^{-1}$. The period of the $I_c(k_f)$ oscillation $\Delta k_f$ is estimated to be $\sim 4.01 \; \mu m^{-1}$, and the corresponding effective cavity length $L_{eff} \sim 0.78 \mu m$.

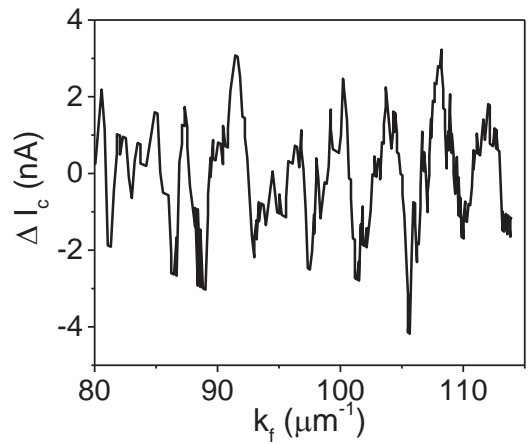

**Supplementary Fig. S3.** The $\Delta I_c$ oscillation as a function of $k_f$. The period $\Delta k_f$ is $\sim 4.01 \; \mu m^{-1}$.

## 4. The Fabry-Pérot oscillations in $G_N$ of Junction-L.

For the long Junction-L, as tuned into the hole conduction regime by gate voltage, the p-n junctions form near the contacts due to the high electron doping near Nb contacts, resulting in the low transmission probability at the interfaces and small $G_N$. The partial reflection of electron waves at the interfaces creates a Fabry-Pérot (F-P) cavity (5, 6). The standing waves lead to pronounced oscillations in conductance as a function of both $V_g$ and applied bias $V_{sd}$. As shown in Fig. S4, an oblique stripe pattern (as marked with dashed lines) is observed, giving a characteristic of Fabry-Pérot oscillations (8, 9). This pattern is only observed under negative gate voltage (n-p-n regime), while it disappears under positive gate voltage (n-$n'$-n regime). This concludes that these oscillations indeed relate to the transmission probability at the interfaces.

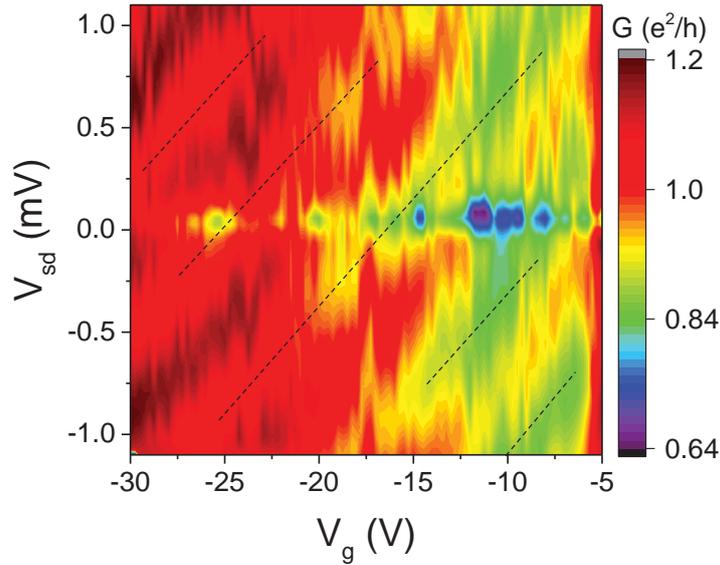

**Supplementary Fig. S4.** The normal state conductance as a function of voltage bias and gate voltage at 12 mK. The dashed lines indicate the oblique stripes, which give a signature of the Fabry-Pérot interference.

## 5. The comparison between $I_c$ and $G_N$ fluctuations as a function of gate voltage.

The comparison between $I_c$ and $G_N$ (taken at $I_{sd} = 30 nA$) fluctuations as a function of gate voltage, measured at 12 mK in Junction-L, is shown in Fig. S5. It is obvious that the patterns of $G_N(V_g)$ and $I_c(V_g)$ fluctuations are not in accordance with each other. This is greatly different with that observed in InAs/InSb nanowire junctions (*10-12*), in which the $G_N$ and $I_c$ fluctuations are almost synchronized. This disagreement in our system can be explained in a regime with different channels between $I_c$ and $G_N$ fluctuations. In such a micrometer long channel of Junction-L, the supercurrent is likely carried by surface states, because the bulk states are nearly de-coherent through the entire nanowire. Thus only the superconducting surface states contribute to the $I_c$ fluctuations, while both surface and bulk states should contribute to the $G_N$ fluctuations.

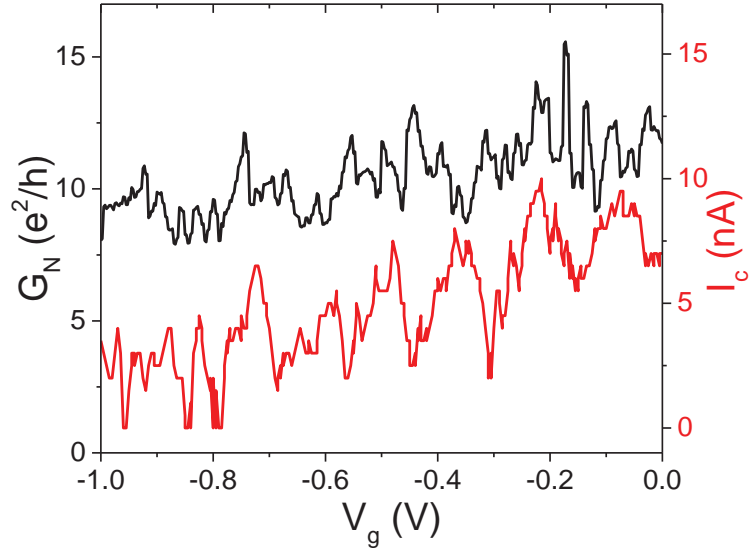

**Supplementary Fig. S5.** Normal state conductance $G_n$ and critical current $I_c$ plotted as a function of gate voltage $V_g$, measured at 12 mK in Junction-L.

# 6. The mobility and mean free path calculated from $G(V_g)$ characteristics.

The carrier mobility and mean free path of the nanowire can be estimated form the $G(V_g)$ characteristics. Figure S6 shows the gate voltage dependence of normal state resistance and conductance of Junction-L, measured at 12 mK with $I_{sd} = 50$ nA in a normal state. From the linear fit (red line) of $G(V_g)$ curve in the region near $V_g = 0$ V, the mobility is calculated to be $\mu_e \sim 2.05 \times 10^4 cm^2/Vs$. The mean free path $l_e = v_f \frac{m^* \mu_e}{e}$ is estimated to be about 467 nm by considering $v_f = 1 \times 10^6 m/s$ and $m^* = 0.04 m_e$. Considering the high electron doping near Nb contacts, below $V_g = 0V$ the hole conduction in the center part and the p-n junction formation near the contacts may also contribute to the sharp increase of resistance. Therefore, the mobility and mean free path may be overestimated. For short Junction-A and B, the heavily doped region from Nb contacts nearly extends to the whole channel, thus the gate modulation capability would be reduced due to screening effect.

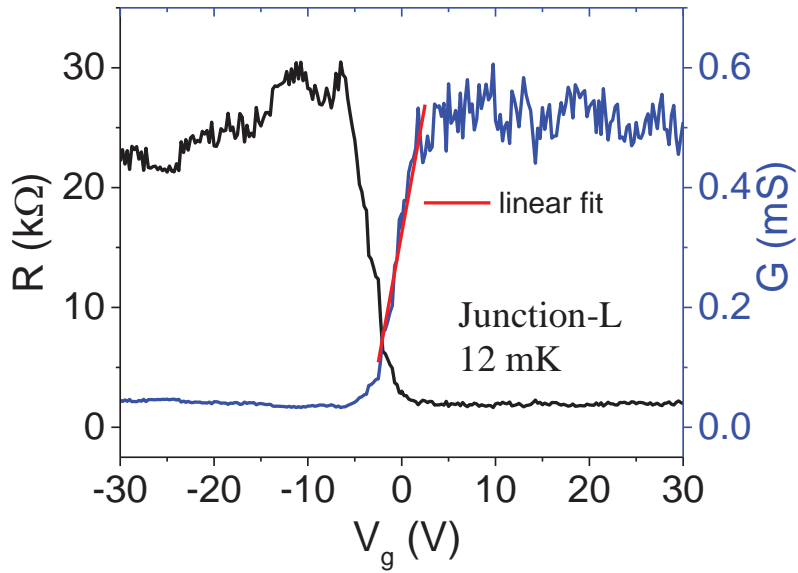

**Supplementary Fig. S6**. Gate voltage dependence of resistance and conductance in Junction-L measured at 12 mK and at $I_{sd} = 50$ nA in a normal state. The red line is the linear fit of $G(V_g)$ curve in the region near $V_g = 0$ V.

## 7. The $I_cR_n$ product of Junction-L

For Junction-A, the critical current $I_c$, normal state resistance $R_n$ and the $I_cR_n$ product as a function of gate voltage are shown in Fig. S7. The $I_cR_n$ is not a constant but varies from 5 to 25 μV. Considering $I_cR_n \leq \frac{\Delta_i}{e}$, the coherence length $\xi = \frac{\hbar v_f}{\pi \Delta_i}$ is estimated to be large than the channel length 1 μm, due to the large Fermi velocity in the range of $v_f \simeq 3 \times 10^5 - 1 \times 10^6 m/s$.

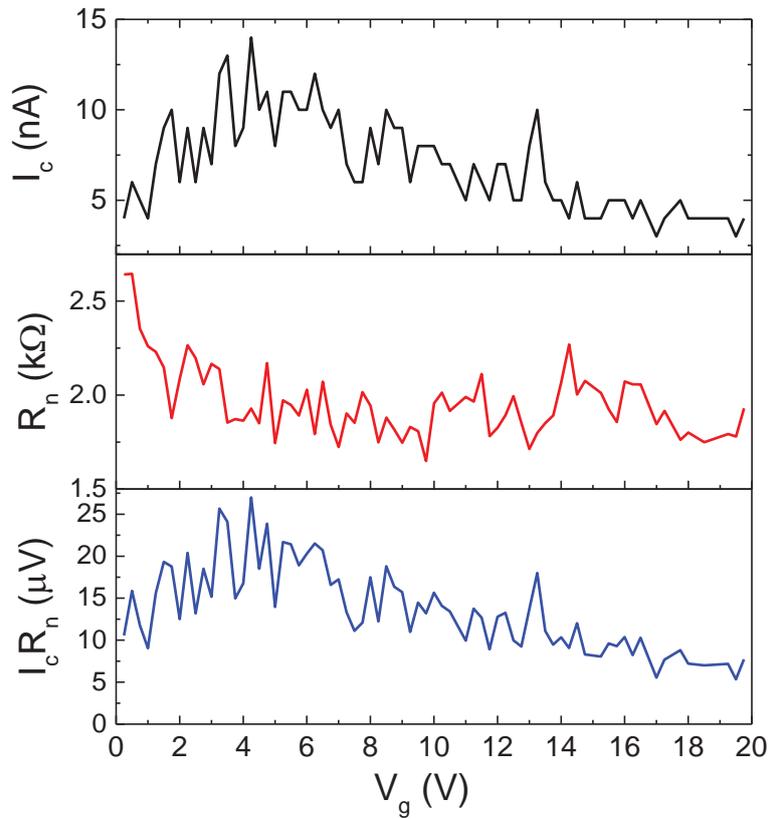

**Supplementary Fig. S7.** The critical current, normal state resistance and their product $I_cR_n$ in Junction-L as a function of gate voltage.